# Некоторые комментарии к статье «Формула для второго вириального коэффициента»

Умирзаков И. Х.

*Новосибирск, Россия*
*e-mail:* [tepliza@academ.org](tepliza@academ.org)

**Аннотация**

Второй вириальный коэффициент, вычисленный с помощью формулы, предложенной в статье Kaplun A.B., Meshalkin A.B. «Equation for the second virial coefficient», опубликованной в High temperature – high pressure, 1999, V. 31, pp. 253-258, сравнивается с экспериментальными данными для гелия-4, водорода, неона, аргона, криптона, ксенона, диоксида углерода, аммиака, воды, метана и этилена. Показано, что эта формула не дает описания в пределах экспериментальной точности экспериментальных данных по зависимости второго вириального коэффициента от температуры всех указанных выше веществ во всем рассмотренном температурном интервале, что противоречит выводам указанной выше статьи. Показано также, что эта формула не дает описания рекомендованных значений второго вириального коэффициента в рамках их точности для гелия-4, водорода, неона, аргона, криптона и метана.

***Ключевые слова***: *второй вириальный коэффициент, водород, гелий-4, неон, аргон, криптон, ксенон, диоксид углерода, вода, аммиак, ртуть, метан, этилена.*

## Введение

Известно [1-3], что $B(T)$ второй вириальный коэффициент (ВВК) однокомпонентного газа определяется через потенциал взаимодействия между двумя одинаковыми частицами (молекулами или атомами), в частности для газов, состоящих из частиц, взаимодействие которых можно считать сферически симметричным, ВВК определяется из

$$B(T) = 2\pi \int_0^\infty [1 - \exp(-U(r)/kT)] r^2 dr,$$

где $U(r)$ - сферически симметричная потенциальная энергия взаимодействия между двумя частицами, расстояние между которыми равно $r$, $T$ – температура, $k$ – постоянная Больцмана.

В работах [4-6] была предложена следующая формула для ВВК

$$B(T) = b + c_2(1 - e^{-\beta/kT}) + c_1(1 - e^{\varepsilon/kT}) - a/T, \qquad (1)$$

где $a, b, c_1, c_2, \beta$ и $\varepsilon$ - неотрицательные параметры. Параметр $a$ можно выразить через температуру Бойля и остальные параметры с помощью

$$a = T_B \cdot [b + c_2(1 - e^{-\beta/kT_B}) + c_1(1 - e^{\varepsilon/kT_B})].$$

В [4] утверждается, что формула (1) описывает экспериментальные данные по ВВК в пределах точности этих данных для гелия-4, водорода, неона, аргона, криптона, ксенона, диоксида углерода, аммиака, воды, ртути, метана и этилена во всем рассматриваемом интервале температур.

В настоящей работе проведено сравнение значений ВВК, вычисленных с помощью формулы (1) со значениями параметров из [4], с экспериментальными данными по ВВК. Показано, что формула (1) не может описывать экспериментальные данные по ВВК указанных выше веществ в пределах погрешности этих данных во всем рассматриваемом интервале температур, если использовать значения параметров формулы (1) из [4]. Это противоречит выводам [4] о том, что обратное имеет место. Показано также, что для части этих веществ формула (1) не описывает даже рекомендованные в [7] значения ВВК в пределах их точности.

**Основная часть**

Ниже проведено сравнение значений ВВК, вычисленных с помощью формулы (1) со значениями параметров из [4], с экспериментальными данными по ВВК из [7-27] и рекомендованными значениями ВВК из [7]. Экспериментальные данные из [8-27] приведены в [7].

Экспериментальные данные на всех рисунках обозначены значками, ошибки экспериментальных [7-27] и рекомендованных данных [7] указаны сплошными линиями. На рисунках $Ta_k$ обозначает значение k-ое значение температуры, $Ba_k$ означает значение ВВК при $Ta_k$, $dB_k$ ($dBpercent_k$) означает ошибку определения экспериментальных или рекомендованных данных. Цифры 1 и 2 после точки в нумерации рисунков означают, что параметры формулы (1) соответственно получены из таблицы 1 и таблицы 2 работы [4], а буква R означает, что точки относятся к рекомендованным данным, если буква R отсутствует в нумерации рисунков, то точки относятся к экспериментальным данным. Источники, откуда взяты данные, указаны в квадратных скобках. Отметим, что в таблицах 1 и 2 в работе [4] параметр $b$ равен нулю.

На рисунках 8.1 и 8.2 экспериментальные данные [18,19] по ВВК диоксида углерода и погрешность определения ВВК, имеющие размерность в $sm^3/g$, переведены в размерность $sm^3/mol$. На всех остальных рисунках ВВК и его погрешность имеет размерность $sm^3/mol$. На рисунках Fig.1.3R и Fig.1.3 использованы параметры $c_1 = -0.17\ sm^3/mol$, $c_2 = 7.937\ sm^3/mol$, $b = 7.307\ sm^3/mol$, $\varepsilon = 5.14\ K$, $\beta = 362.766\ K$, $T_B = 25\ K$. На рисунках Fig.1.4R и Fig.1.4 использованы параметры $c_1 = 0$, $c_2 = 7.929\ sm^3/mol$, $b = 7.24\ sm^3/mol$, $\beta = 372.078\ K$, $T_B = 25\ K$.

**Гелий-4  He.**

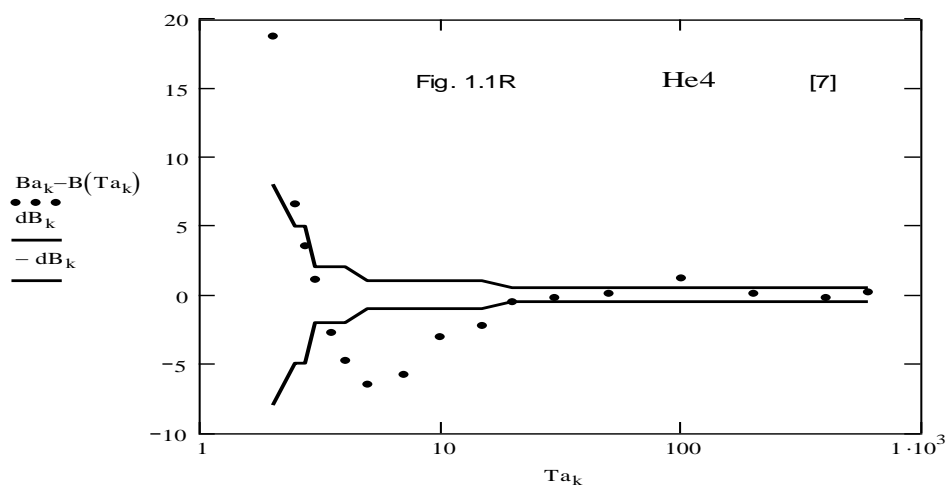

Fig. 1.2R  He4  [7]

$Ba_k - B(Ta_k)$
$dB_k$
$-dB_k$

$Ta_k$

Fig. 1.3R  He4  [7]

$Ba_k - B(Ta_k)$
$dB_k$
$-dB_k$

$Ta_k$

Fig. 1.4R  He4  [7]

$Ba_k - B(Ta_k)$
$dB_k$
$-dB_k$

$Ta_k$

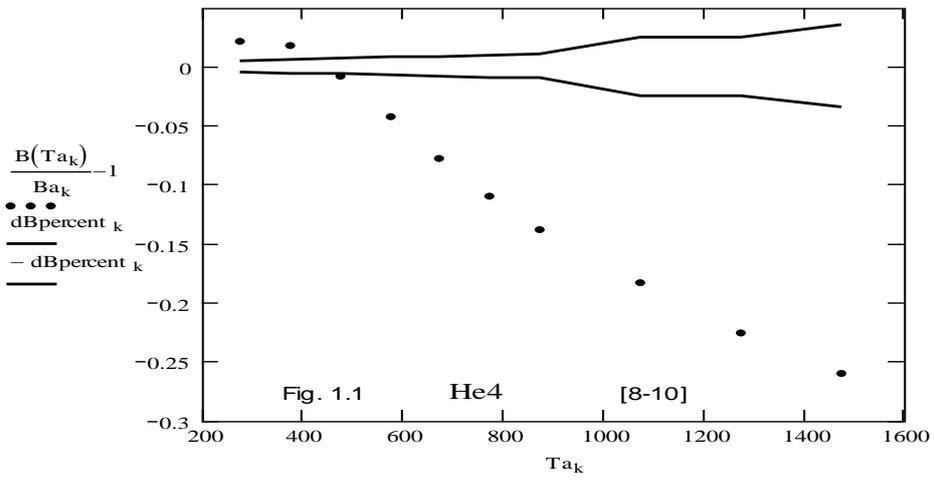

Fig. 1.1   He4   [8-10]

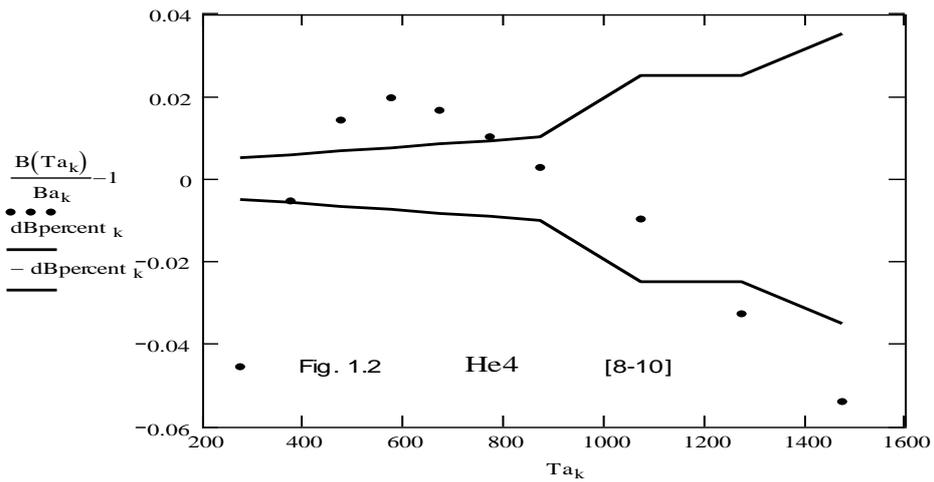

Fig. 1.2   He4   [8-10]

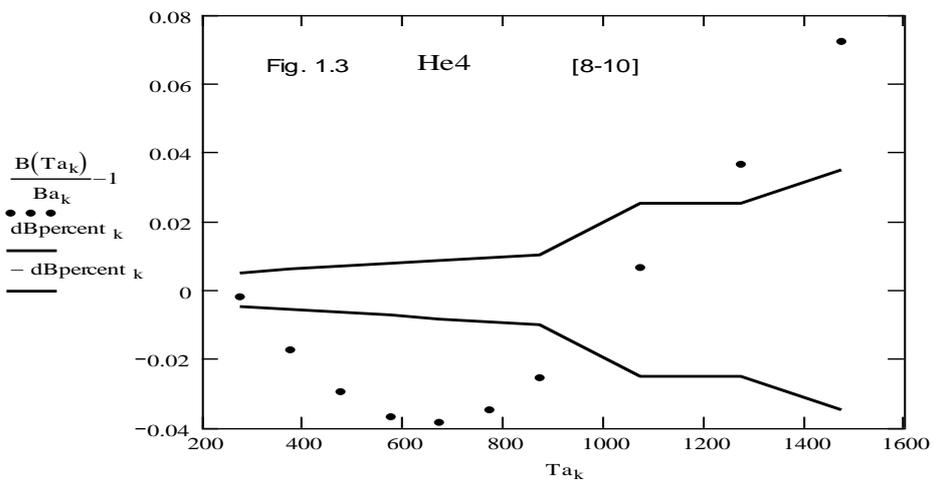

Fig. 1.3   He4   [8-10]

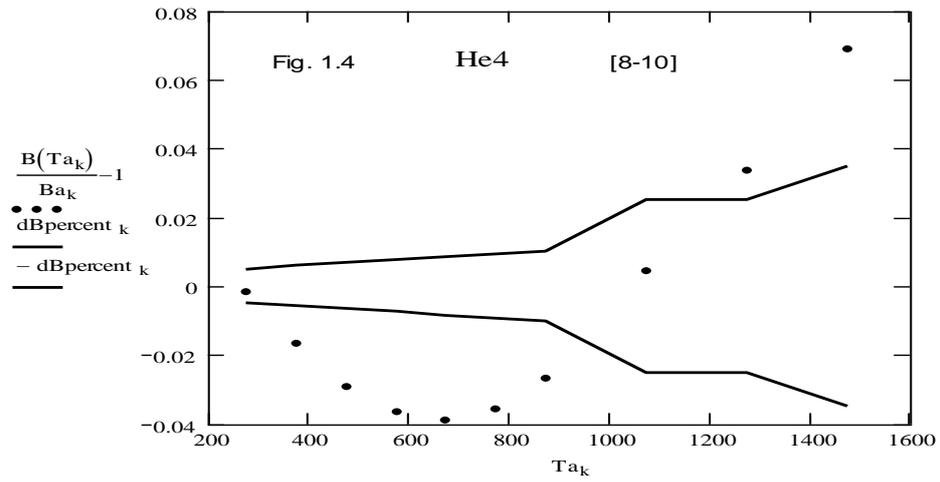

## Молекулярный водород $H_2$.

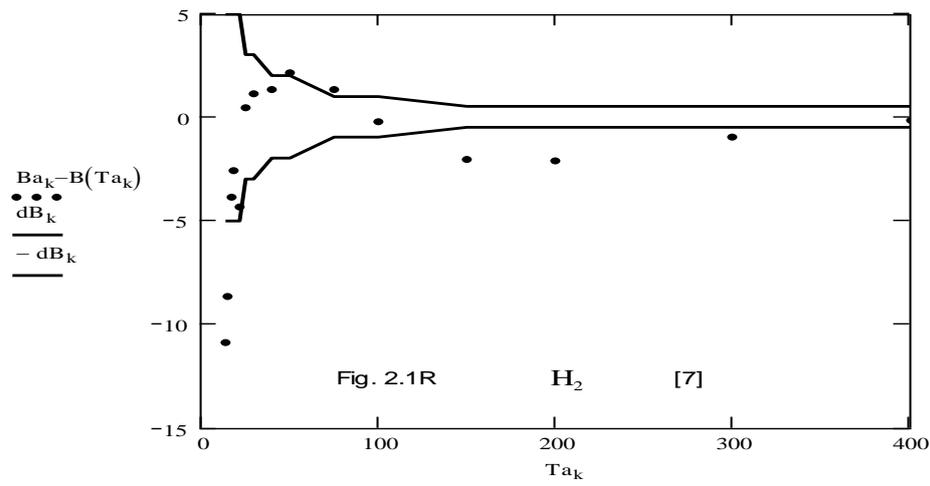

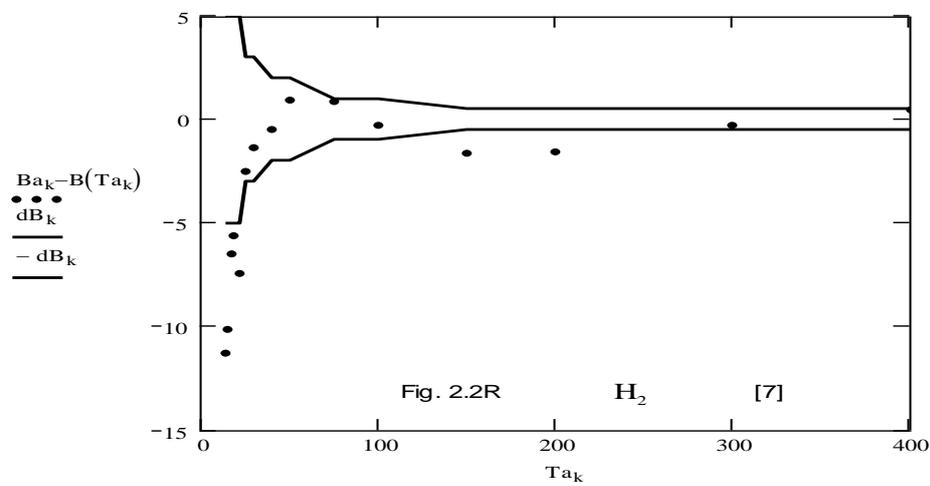

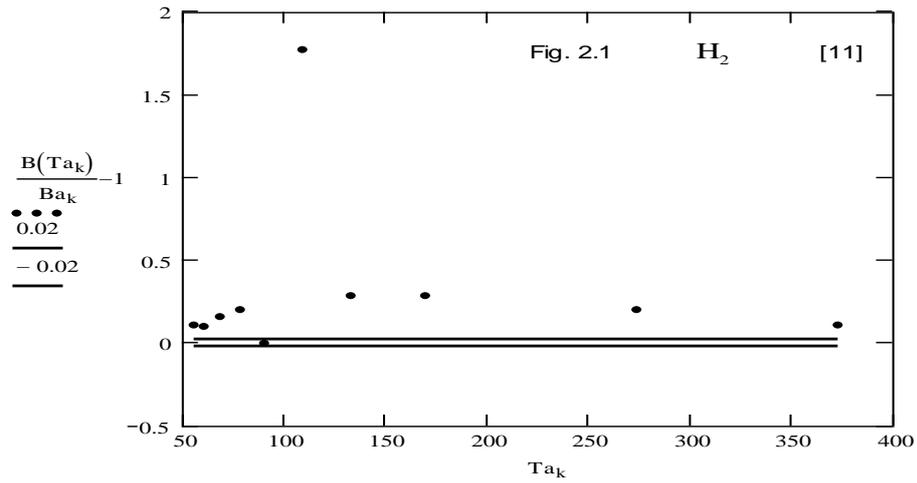

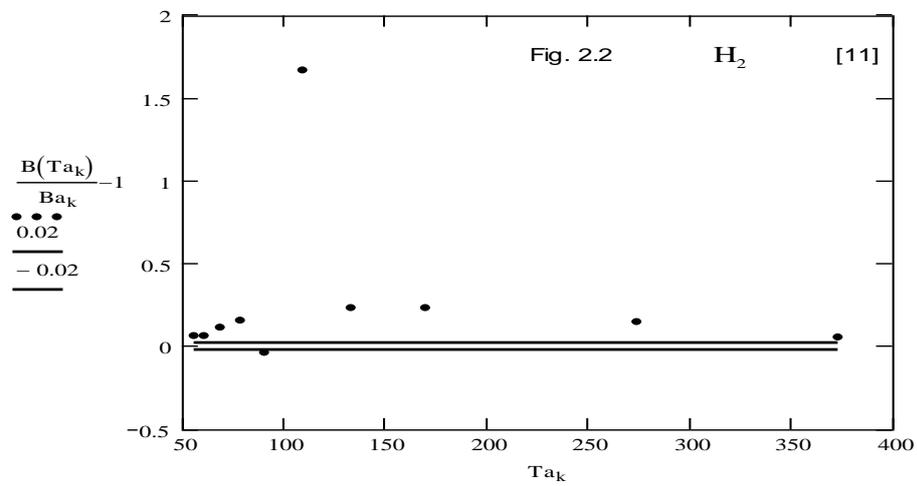

## Неон Ne.

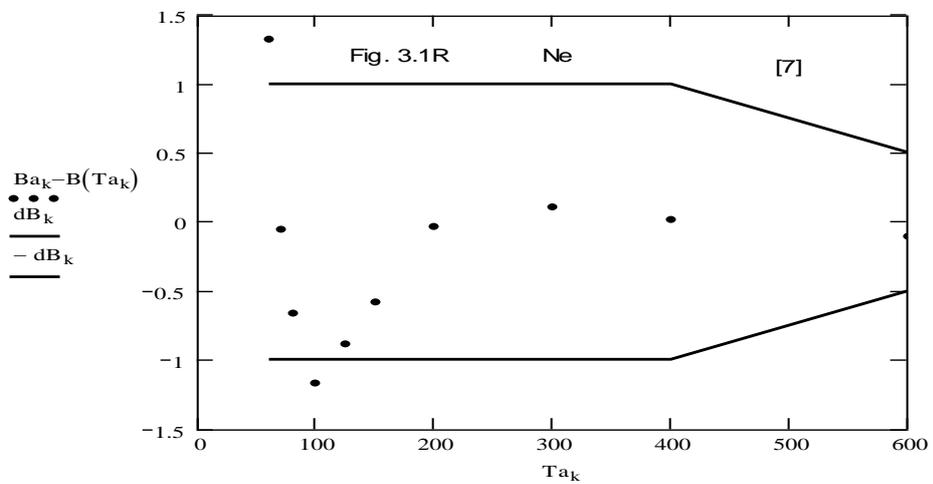

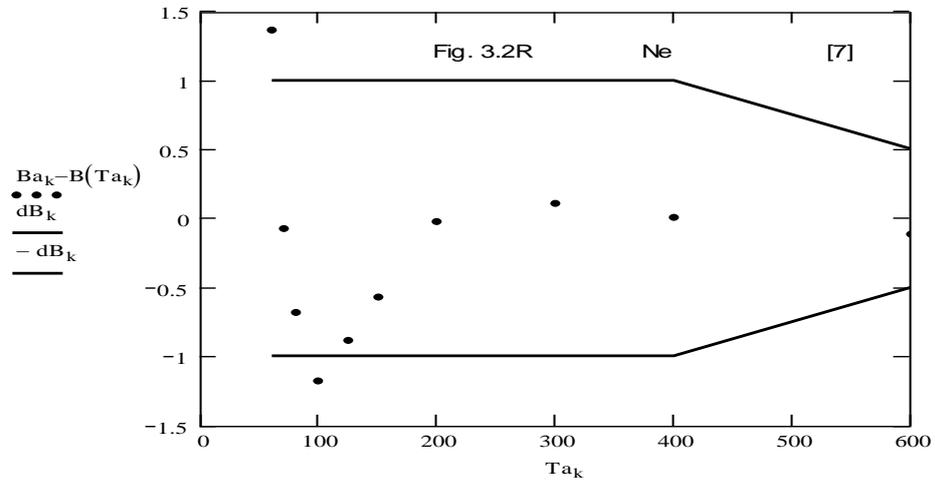

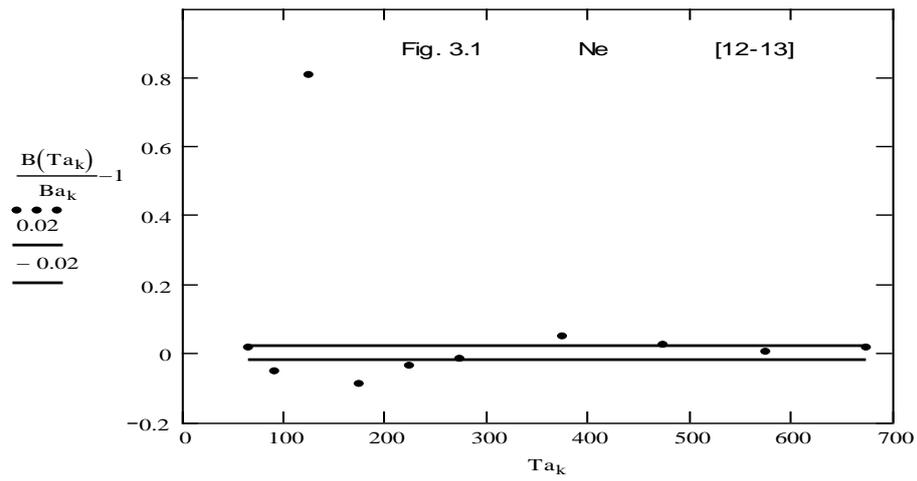

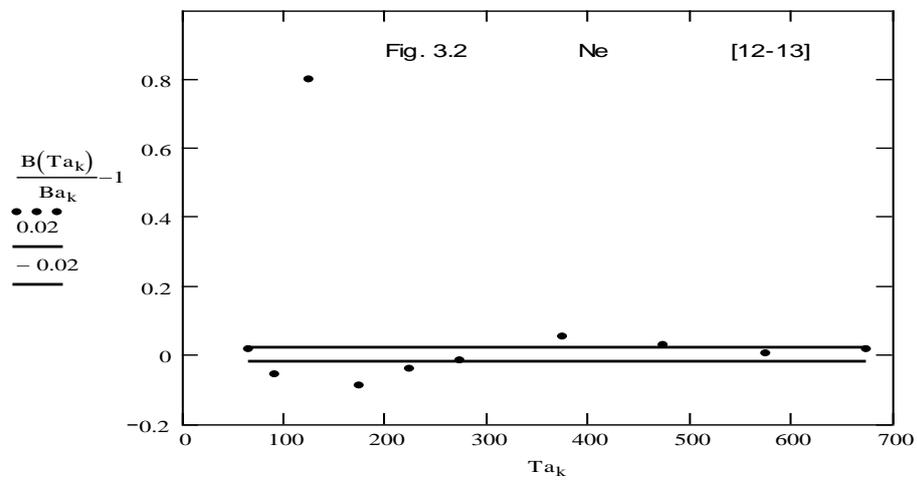

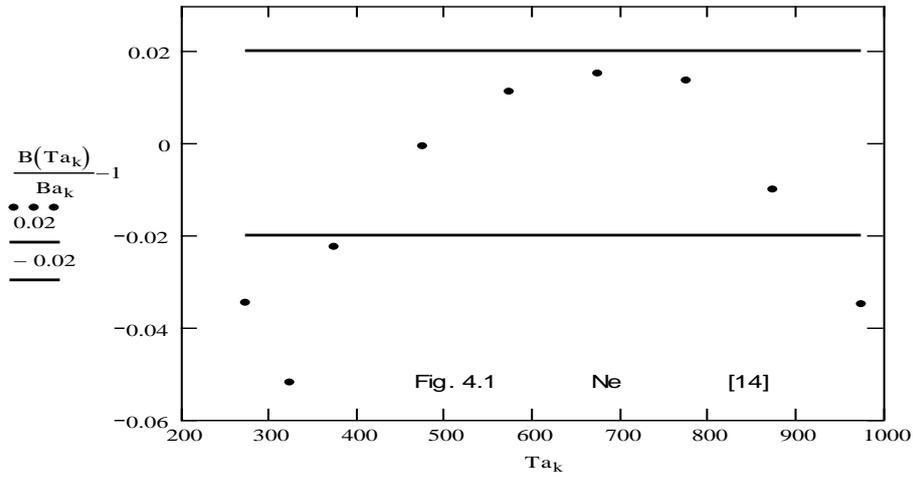

Fig. 4.1 Ne [14]

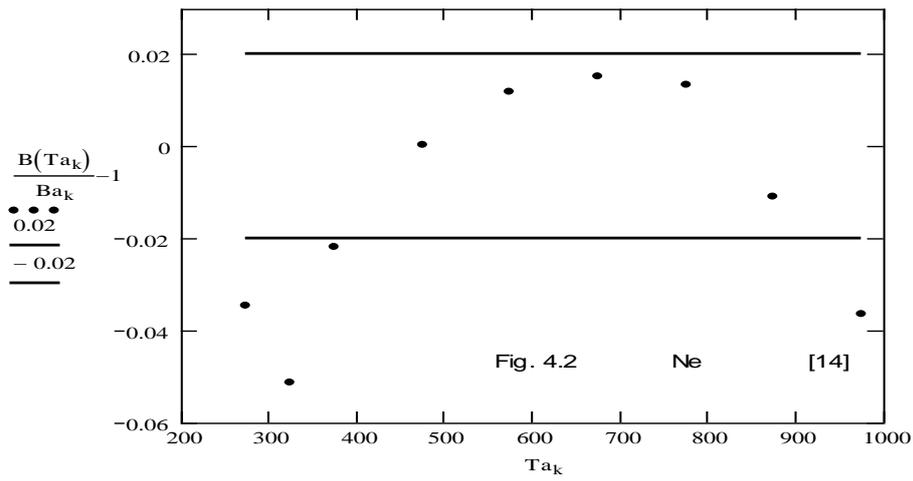

Fig. 4.2 Ne [14]

**Аргон Ar.**

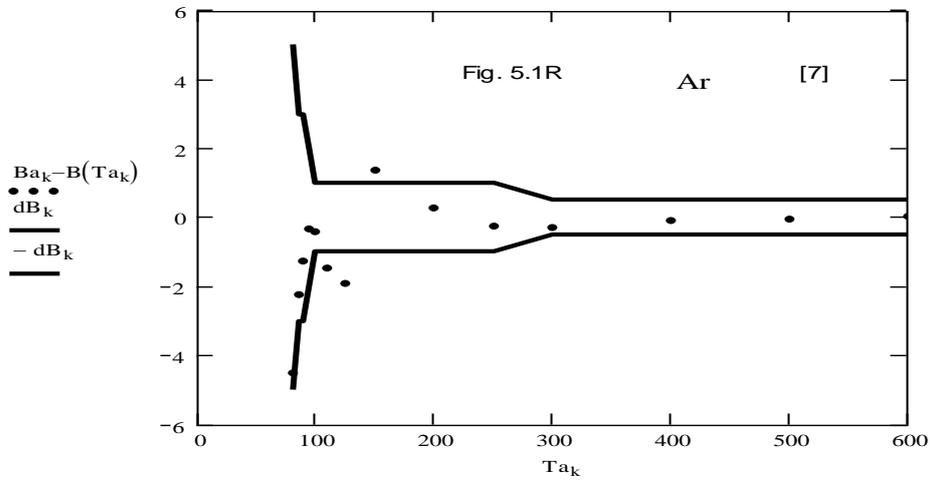

Fig. 5.1R Ar [7]

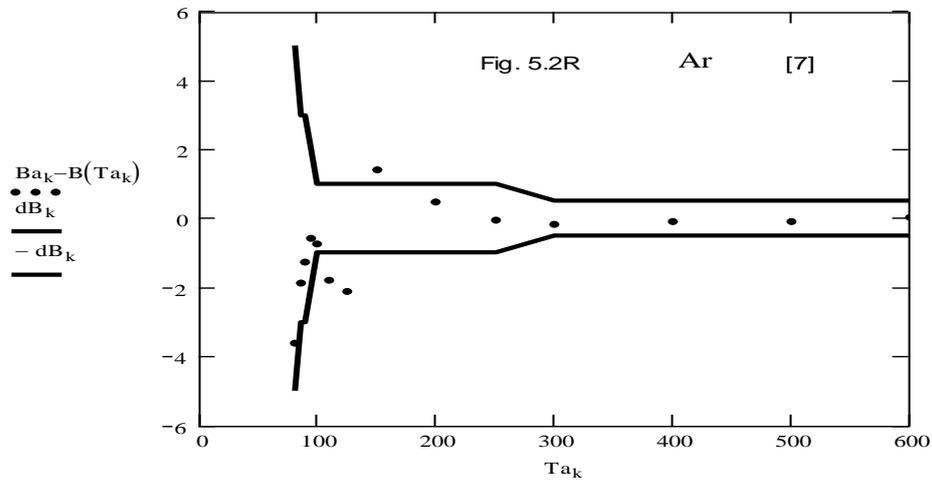
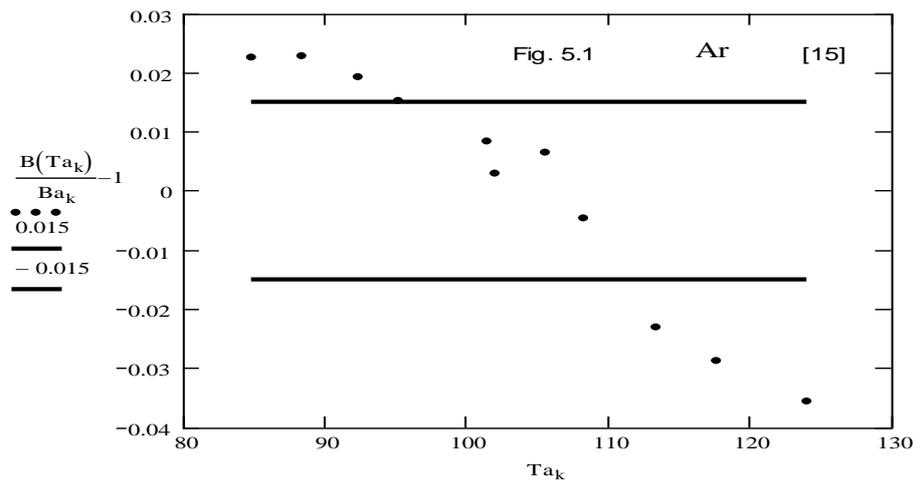
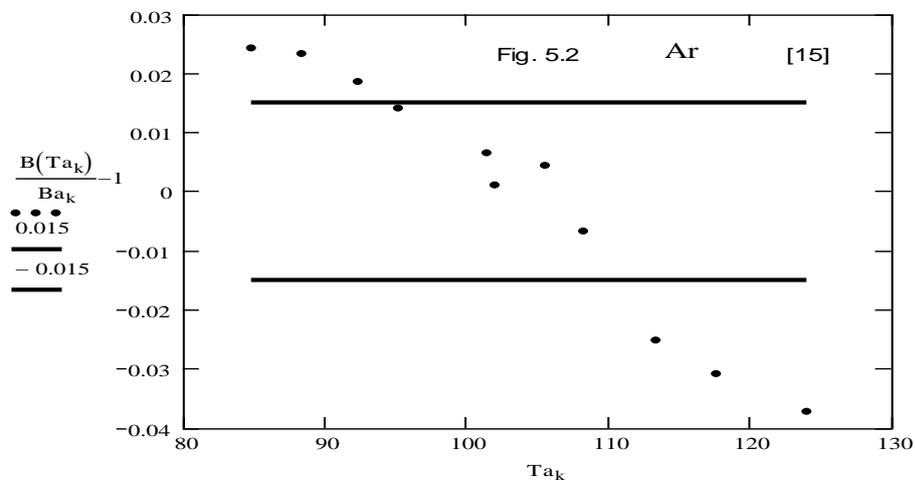

# Криптон Kr.

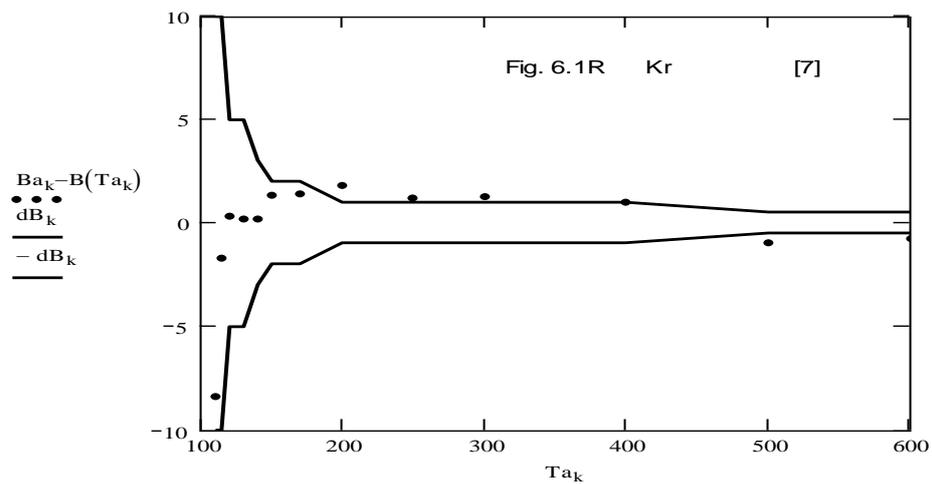

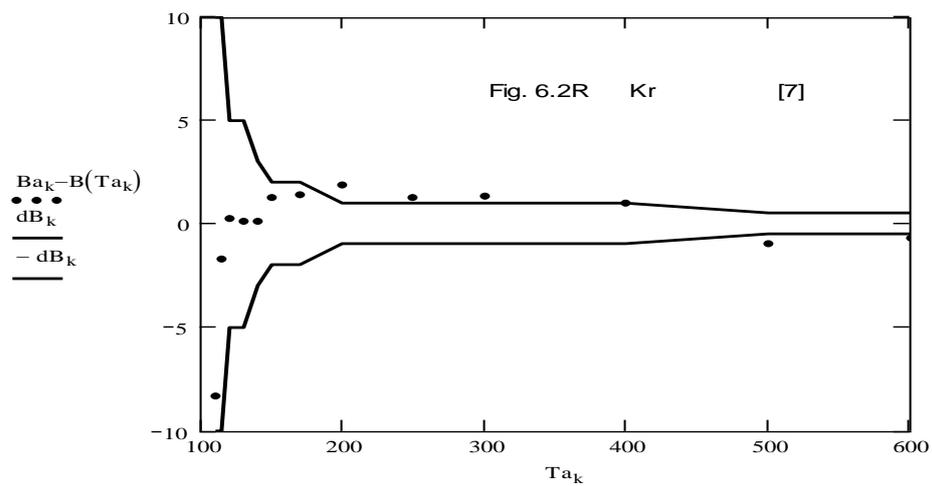

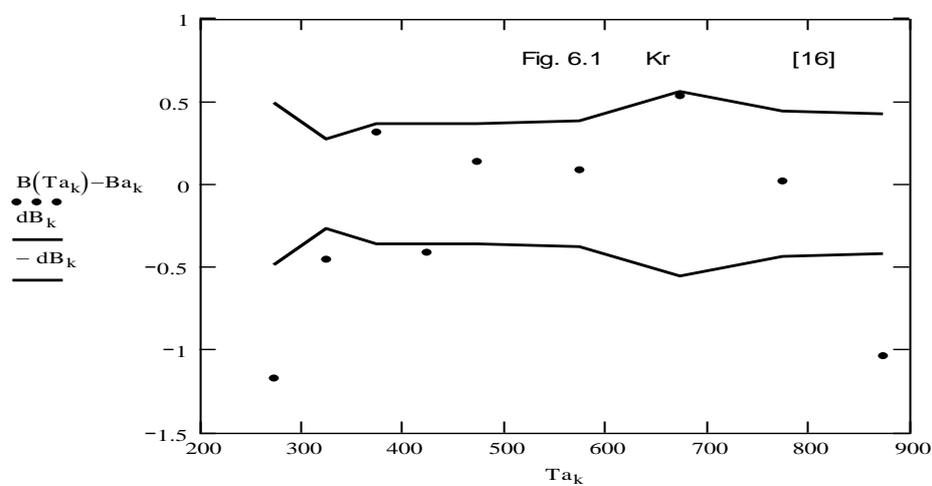

Fig. 6.2 Kr [16]

**Ксенон Xe.**

Fig. 7.1 Xe [17]

Fig. 7.2 Xe [17]

**Диоксид углерода $CO_2$.**

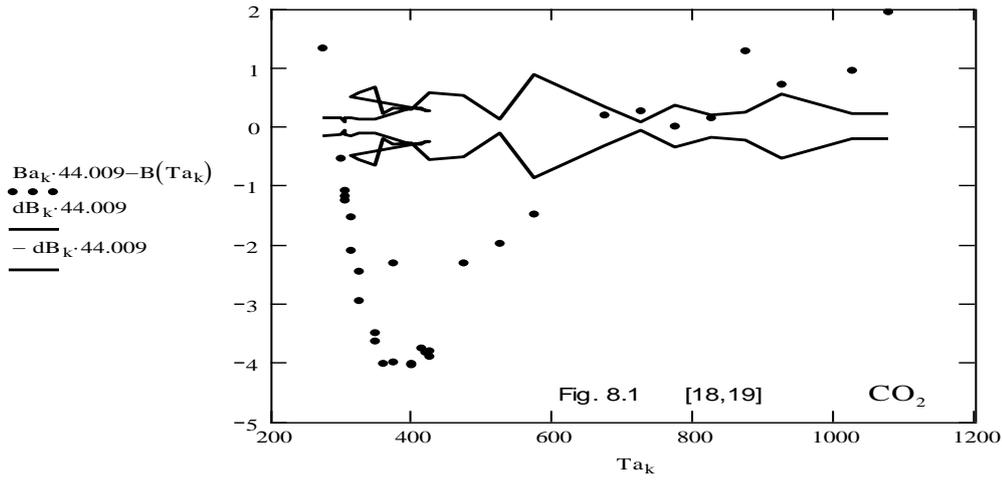

Fig. 8.1 [18,19] $CO_2$

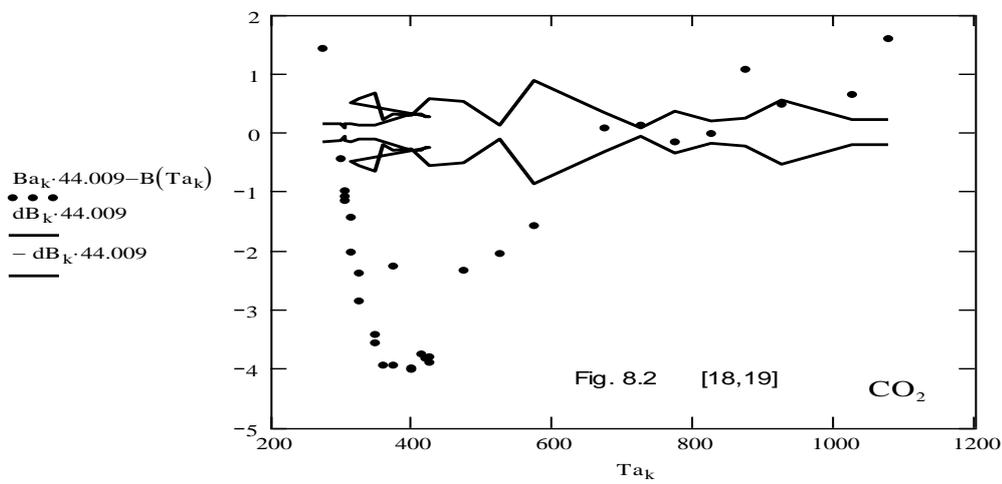

Fig. 8.2 [18,19] $CO_2$

**Вода $H_2O$.**

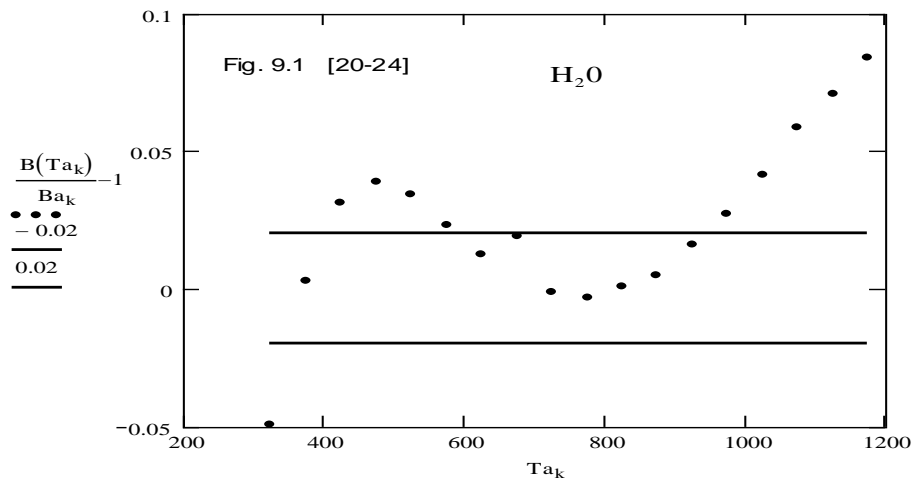

Fig. 9.1 [20-24] $H_2O$

**Метан CH$_4$.**

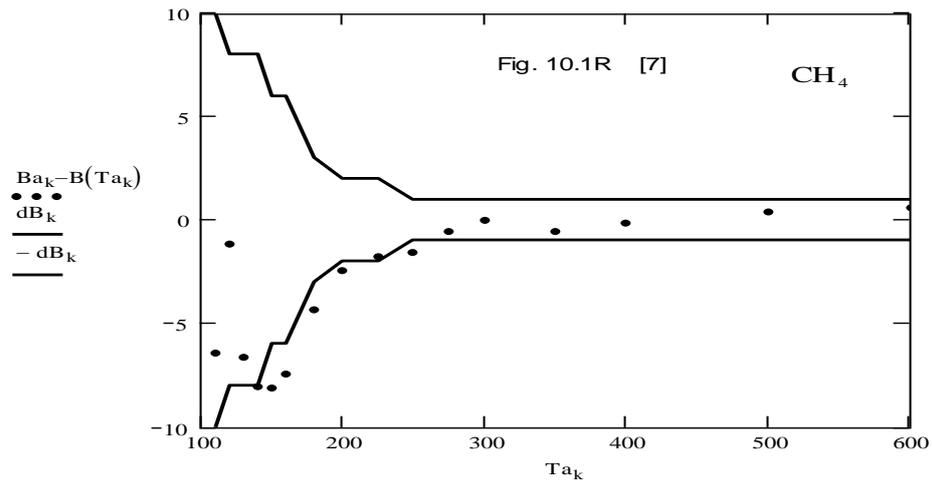

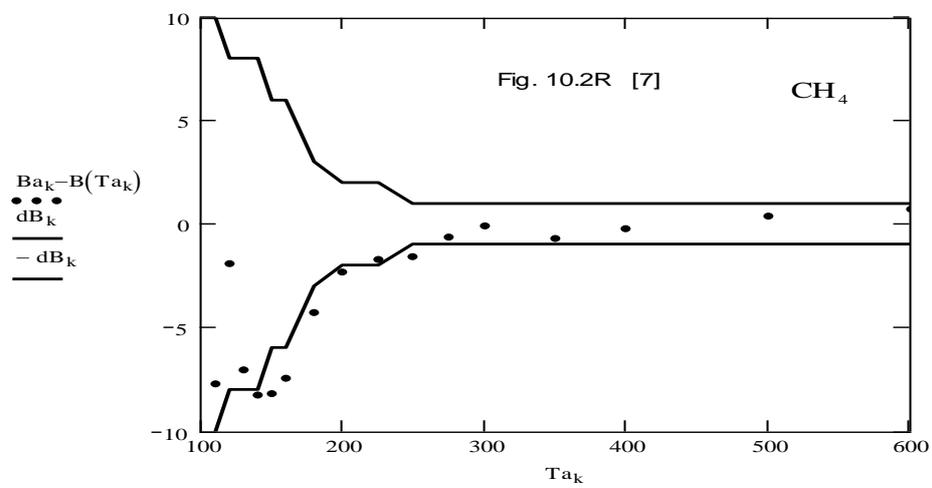

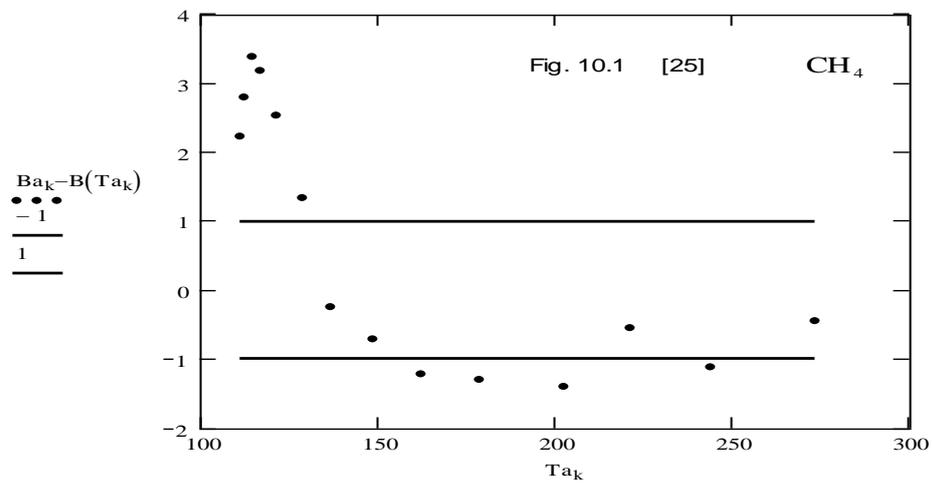

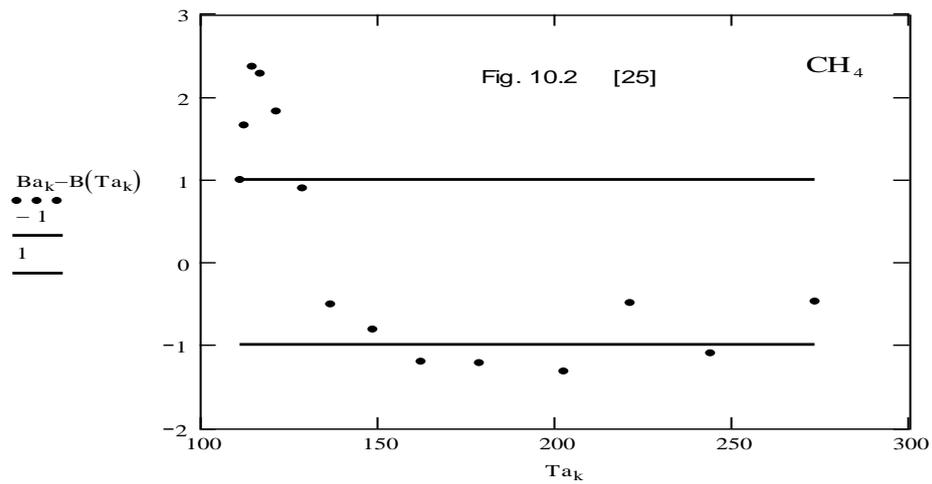

## Этилен C$_2$H$_4$.

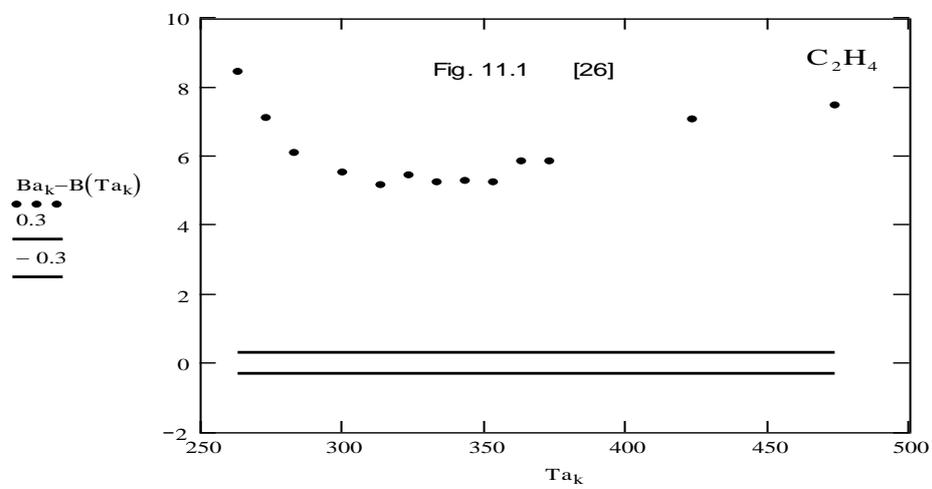

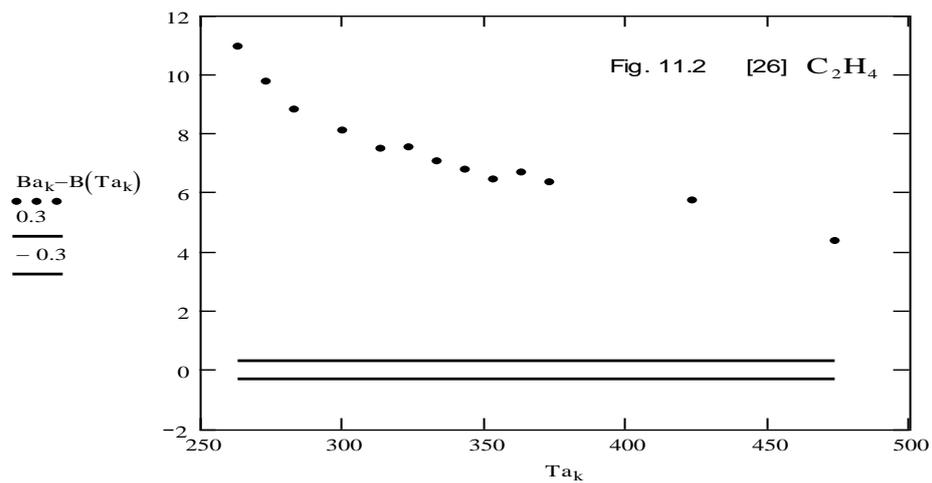

**Аммиак $NH_3$.**

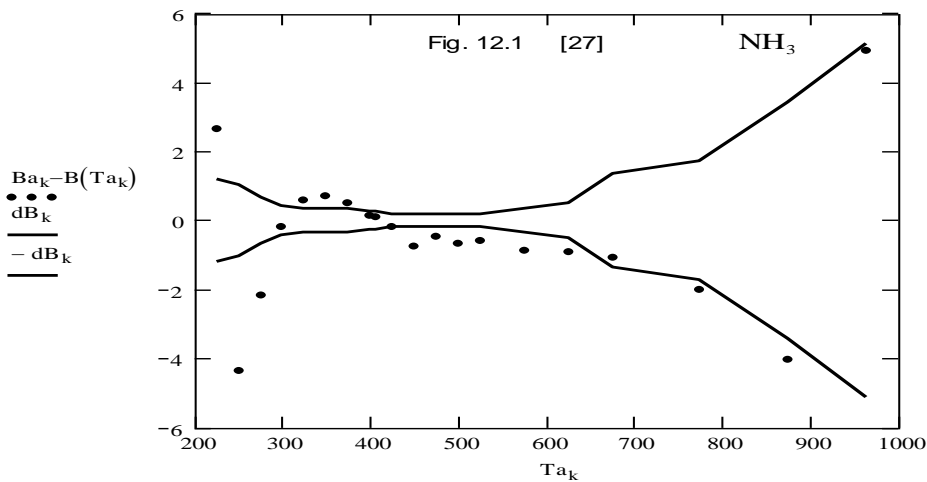

**Ртуть Hg.**

Нами не обнаружено ни в одной из цитированных в [4] работах экспериментальных данных по ртути с указанием их точности, поэтому неизвестно, на основе чего в [4] сделан вывод о том, что формула (1) описывает экспериментальные данные по ртути с экспериментальной точностью.

## Результаты и их обсуждение

Как видно из рисунков, формула (1) не описывает экспериментальные данные по второму вириальному коэффициенту гелия-4, водорода, неона, аргона, криптона, ксенона, диоксида углерода, воды, аммиака, метана и этилена с экспериментальной точностью во всем рассмотренном интервале температур.

Из рисунков также видно, что формула (1) не описывает даже рекомендованные в [7] данные по ВВК с их точностью определения для гелия-4, водорода, неона, аргона, критона и метана.

Ошибки определения ВВК с помощью формулы (1) больше в разы, а для некоторых веществ в десятки раз больше, чем ошибки (погрешности) экспериментальных и рекомендованных данных по ВВК. Следовательно, выводы работы [4] о том, что имеет место обратное, неверно.

В работах [5-6] утверждается, что предложенный в них потенциал взаимодействия описывает экспериментальные данные по второму вириальному коэффициенту многих веществ в рамках экспериментальной точности. Проверка, проведенная нами, показала, что это не имеет места для всех рассмотренных в [4-6] веществ. Отличие по ВВК для гелия -4 при этом составляет не десятки процентов, даже не сотни процентов, а в 1200 раз при низких температурах. Результаты этой проверки составляет предмет следующей публикации.

Обнаружено также, что авторы работ [4-5] в ряде своих публикаций (смотрите, например, [28,29]) уравнение состояния, принадлежащее ван Лаару, выдают как свое собственное. Более того, они в своих работах об этом уравнении состояния ссылаются на источник [30], где подробно разбирается уравнение состояния ван Лаара. Налицо явный неприкрытый плагиат.

Кроме того, имеются серьезные физические ошибки в теории вибрационного метода измерения вязкости, предложенной первым из авторов [4], в самом вибрационном методе измерения вязкости жидкостей, а также в методе вибрационного фазового анализа одно- и многокомпонентных систем при их плавлении и затвердевании, основанных на указанной теории. Кроме этого, эти методы базируются на физически невыполнимых предположениях. Поэтому результаты, полученные на основе этих методов, являются ошибочными. На этих методах и на результатах, полученных этими методами, основаны кандидатские и докторские диссертации, защищенные этими авторами.

Следовательно, есть большие сомнения в добросовестности авторов [4] при проведении научных исследований, и ни одну из публикаций этих авторов нельзя принимать на веру без серьезной проверки каждого положения и экспериментальных результатов в них, лучше их вообще не использовать.

До сих пор остается открытым вопрос применимости указанных методов для физико-химических исследований, так как не проведена проверка этих методов независимыми исследователями.

**Выводы**

Формула для второго вириального коэффициента, предложенная в работе Kaplun A.B., Meshalkin A.B. «Equation for the second virial coefficient», опубликованной в High temperature – high pressure, 1999, V. 31, pp. 253-258, не может описывать экспериментальные данные по ВВК в пределах погрешности этих данных во всем рассматриваемом интервале температур для гелия-4, водорода, неона, аргона, криптона, ксенона, диоксида углерода, аммиака, воды, метана, этилена, Это противоречит выводам этой работы, где утверждается обратное.

Показано, что формула (1) не описывает даже рекомендованные в [7] значения ВВК в пределах их точности для гелия-4, водорода, неона, аргона, критона и метана.

**Литература**

# Some comments on «Equation for the second virial coefficient»

Umirzakov I. H.

*Novosibirsk, Russia*
*e-mail: tepliza@academ.org*

**Abstract**

The second viral coefficient calculated using the equation suggested in the paper of Kaplun A.B., Meshalkin A.B. Equation for the second virial coefficient published in High temperature high pressure, 1999, Volume 31, pages 253-258 is compared with experimental data for helium, hydrogen, neon, argon, krypton, xenon, carbon dioxide, water, ammonia, methane and ethylene. It is shown the formula cannot describe the temperature dependence of the experimental data on the second virial coefficient for the all above substances within the experimental error over the investigated temperature interval. The latter is in controversy with the derivations of the paper mentioned above. It is also shown the formula cannot describe the recommended data for the second virial coefficient within their uncertainties for helium, hydrogen, neon, argon, krypton and methane.

***Keywords***: *second viral coefficient, helium-4, hydrogen, neon, argon, krypton, xenon, carbon dioxide, water, mercury, ammonia, methane, ethylene.*